\documentclass[preprint,showpacs,showkeys,preprintnumbers,amsmath,amssymb]
{revtex4}

\usepackage{graphicx}
\usepackage{dcolumn}
\usepackage{bm}

\begin{document}


\title{Searching for Decoherence-Free Subspaces in Cavity Quantum
Electrodynamics}  

\author{A. R. Bosco de Magalh\~aes}
\email{arbm@fisica.ufmg.br}
\affiliation{Departamento de F\'{\i}sica, ICEX, Universidade Federal de
Minas Gerais, C.P. 702, 30161-970 Belo Horizonte, MG, Brazil}

\author{M. C. Nemes}
\email{carolina@fisica.ufmg.br}
\affiliation{Departamento de F\'{\i}sica, ICEX, Universidade Federal
de Minas Gerais, C.P. 702, 30161-970 Belo Horizonte, MG, Brazil}

\date{\today}

\begin{abstract}
We construct a model which describes a recently performed experiment (%
\textit{Phys. Rev. A} \textbf{64}, 050301(R) (2001)) in which an entangled
state between two modes of a single cavity is built. Environmental effects
are taken into account and the results agree with the experimental findings.
Moreover the model predicts, for different conditions of the same
experiment, a decoherence-free subspace. These conditions are analyzed and
slightly different experiments suggested in order to test its viability.
\end{abstract}



\maketitle

\section{Introduction}

The role of environmental degrees of freedom is a central theme today both
in the study of Foundations of Quantum Mechanics \cite{art1} and for technological
improvement in what concerns Quantum Computation \cite{art2}. Cavity Quantum
Electrodynamics \cite{art3} is one of the areas where such problem has been
intensively investigated in order to establish the dynamics of the evolution
of quantum superpositions considered as an open system.

Decoherence control is an important topic in this context. Systems composed
by two parts interacting with a common reservoir are of interest, since
theoretically they may lead to the existence of trapped states. This effect
shall be associated to cross decay rate terms, which have been studied at
least since Agarwall \cite{art4}. Derivations of master equations for such systems
may be found in \cite{art5,art6,art7,art8,art9} 
and applications of them in \cite{art10,art11,art12,art13}.

A recent and particularly interesting experiment involves the construction
of two electromagnetic field modes in a single cavity \cite{art14}. In the present
contribution we investigate the consequences of a straightforward
generalization of the Caldeira-Legget type model for the discussed
experiment and are able to explain the results. Moreover this model predicts
the existence of a decoherence-free subspace (DFS) provided the coupling to
the environment satisfies certain conditions. These cannot be tested in the
present experiment although with slight modifications the feasibility and
robustness of such spaces can be easily accessed.

In Sec. II we present a Hamiltonian to model the system (two electromagnetic
field modes) subjected to the environment, and derive a master equation in
the usual Markov regime. A technique for solving this master equation, group
theory for superoperators, is given in Appendix A. The experiment is
described in Sec. III, where its results are compared to the theoretical
ones from our model. In Sec. IV we suggest slight modifications in the
experimental sketch, useful to investigate a possible tendency of forming
DFS in such systems. Our conclusions are presented in Sec. V, where the
feasibility of DFS is analyzed.

\section{From the model to the master equation}

The Hamiltonian we use to describe two electromagnetic field modes in a
single superconducting cavity plus environment is

\begin{equation}
\mathbf{H}=\mathbf{H}_{0}+\mathbf{H}_{int},  \label{H}
\end{equation}
where 
\begin{eqnarray}
\mathbf{H}_{0} &=&\mathbf{H}_{S}+\mathbf{H}_{E},  \label{Hamiltoniano} \\
\mathbf{H}_{S} &=&\hbar \Omega _{1}\mathbf{a}_{1}^{\dagger }\mathbf{a}%
_{1}+\hbar \Omega _{2}\mathbf{a}_{2}^{\dagger }\mathbf{a}_{2},  \notag \\
\mathbf{H}_{E} &=&\hbar \underset{k}{\sum }\omega _{k}\mathbf{c}%
_{k}^{\dagger }\mathbf{c}_{k},  \notag \\
\mathbf{H}_{int} &=&\underset{k}{\hbar \sum }\left( \alpha _{1k}\mathbf{a}%
_{1}\mathbf{c}_{k}^{\dagger }+\alpha _{1k}^{\ast }\mathbf{a}_{1}^{\dagger }%
\mathbf{c}_{k}\right) +\hbar \underset{k}{\sum }\left( \alpha _{2k}\mathbf{a}%
_{2}\mathbf{c}_{k}^{\dagger }+\alpha _{2k}^{\ast }\mathbf{a}_{2}^{\dagger }%
\mathbf{c}_{k}\right) .  \notag
\end{eqnarray}
The operators $\mathbf{a}_{1}$ ($\mathbf{a}_{2}$) and $\mathbf{a}%
_{1}^{\dagger }$ ($\mathbf{a}_{2}^{\dagger }$) are annihilation and creation
bosonic operators for mode $M_{a}$ ($M_{b}$) with frequency $\Omega _{1}$ ($%
\Omega _{2}$). The environment is modelled by a set of harmonic oscillators
with creation and annihilation operators $\mathbf{c}_{k}^{\dagger }$ and $%
\mathbf{c}_{k}$, linearly coupled to the system, as it is usually 
done \cite{art15}.
Harmonic oscillators are appropriate to model extended modes \cite{art16}, as
phonons in the cavity mirrors and electromagnetic environment modes in the
laboratory. The coupling between these environment oscillators and the
oscillators of interest may occur by complicated processes \cite{art17} 
(\emph{e.g.}%
, a photon may be scattered from $M_{a}$ to an electromagnetic environment
mode by an spurious atom inside the cavity). The coupling we considered is
an effective one, related to changes of one photon between environment and
system.

In what follows we deduce a master equation for the system in a similar
fashion as done in Ref. \cite{art9}, where we also considered a coupling between the
relevant modes (more detailed calculation can be found in Ref. \cite{art9}). Let us
take $\mathbf{\rho }\left( t\right) $ the complete density operator
concerning the system plus environment. Its time evolution may be given by 
\begin{equation}
\frac{d}{dt}\mathbf{\tilde{\rho}}\left( t\right) =\frac{-i}{\hbar }\left[ 
\tilde{\mathbf{H}}_{int}\left( t\right) ,\mathbf{\tilde{\rho}}\left(
t\right) \right] ,  \label{vonNewman}
\end{equation}
where 
\begin{equation*}
\mathbf{\tilde{\rho}}\left( t\right) =e^{\frac{i}{\hbar }\mathbf{H}_{0}t}%
\mathbf{\rho }\left( t\right) e^{-\frac{i}{\hbar }\mathbf{H}_{0}t},\text{%
\quad \quad }e^{\frac{i}{\hbar }\mathbf{H}_{0}t}\mathbf{H}_{int}\left(
t\right) e^{-\frac{i}{\hbar }\mathbf{H}_{0}t}
\end{equation*}
are in the interaction picture. The high quality factor of the cavity permit
us to consider $\tilde{\mathbf{H}}_{int}\left( t\right) $ small, since $%
M_{a} $ and $M_{b}$ are weakly coupled to the environment. Disregarding the
terms of third order in $\tilde{\mathbf{H}}_{int}\left( t\right) $, Eq. (\ref
{vonNewman}) leads to 
\begin{equation}
\mathbf{\tilde{\rho}}\left( t\right) -\mathbf{\tilde{\rho}}\left( 0\right) =-%
\frac{i}{\hbar }{\int_{0}^{t}}dt^{\prime }[\mathbf{\tilde{H}}_{int}\left(
t^{\prime }\right) ,\mathbf{\tilde{\rho}}\left( 0\right) ]-\frac{1}{\hbar
^{2}}{\int_{0}^{t}}dt^{\prime }{\int_{0}^{t^{\prime }}}dt^{\prime \prime }[%
\mathbf{\tilde{H}}_{int}\left( t^{\prime }\right) ,[\mathbf{\tilde{H}}%
_{int}\left( t^{\prime \prime }\right) ,\mathbf{\tilde{\rho}}\left( 0\right)
]].  \label{integr}
\end{equation}

Let us admit that at $t=0$ the system is prepared in the state $\mathbf{\rho 
}_{S}\left( 0\right) $ and the environment is in thermal equilibrium. Thus 
\begin{equation}
\mathbf{\tilde{\rho}}\left( 0\right) =\mathbf{\rho }_{S}\left( 0\right)
\otimes \mathbf{\rho }_{E}\left( 0\right) ,
\end{equation}
with 
\begin{equation}
\mathbf{\rho }_{E}\left( 0\right) =\frac{1}{Z}{\prod_{k}}\exp \left( -\beta
\hbar \omega _{k}\mathbf{c}_{k}^{\dagger }\mathbf{c}_{k}\right) ,\text{\quad
\quad }Z={\prod_{k}}\sum_{n=0}^{\infty }\exp \left( -\beta \hbar \omega
_{k}n\right) ,\text{\quad \quad }\beta =\left( k_{B}\mathcal{T}\right) ^{-1}.
\end{equation}
Here $k_{B}$ is Boltzmann constant, $\mathcal{T}$ is the absolute
temperature and $Z$ is the partition function \cite{art18}. Taking the trace over
the environment degrees of freedom in both sides of Eq. (\ref{integr}), we
can find, in the limit of zero temperature, 
\begin{eqnarray}
\mathbf{\tilde{\rho}}_{S}\left( t\right) -\mathbf{\tilde{\rho}}_{S}\left(
0\right)  &=&\sum_{i,j=1}^{2}{\int_{0}^{t}}dt^{\prime }{\int_{0}^{t^{\prime }%
}}d\tau \left\{ \left( \sum_{k}\alpha _{ik}\alpha _{jk}^{\ast }e^{i\omega
_{k}\tau }\right) e^{-i\omega _{j}\tau }e^{i\left( \Omega _{j}-\Omega
_{i}\right) t^{\prime }}\left( \mathbf{a}_{i}\mathbf{\tilde{\rho}}_{S}\left(
0\right) \mathbf{a}_{j}^{\mathbf{\dagger }}-\mathbf{\tilde{\rho}}_{S}\left(
0\right) \mathbf{a}_{j}^{\mathbf{\dagger }}\mathbf{a}_{i}\right) \right\}  
\notag \\
&&+\text{h.c.},
\end{eqnarray}
where h.c. stands for Hermitian conjugate.

Notice that $\sum_{k}\alpha _{ik}\alpha _{jk}^{\ast }e^{i\omega _{k}\tau }$
decays very fast with the growing of $\tau $. Thus we may modify the
integration limits above and obtain 
\begin{equation}
\mathbf{\tilde{\rho}}_{S}\left( t\right) -\mathbf{\tilde{\rho}}_{S}\left(
0\right) =\sum_{i,j=1}^{2}\left\{ 
\begin{array}{c}
\left( k_{ij}+i\Delta _{ij}\right) \left( \mathbf{a}_{i}\mathbf{\tilde{\rho}}%
_{S}\left( 0\right) \mathbf{a}_{j}^{\mathbf{\dagger }}-\mathbf{\tilde{\rho}}%
_{S}\left( 0\right) \mathbf{a}_{j}^{\mathbf{\dagger }}\mathbf{a}_{i}\right)
\left( {\int_{0}^{t}}dt^{\prime }e^{i\left( \omega _{j}-\omega _{i}\right)
t^{\prime }}\right) 
\end{array}
\right\} +\text{h.c.}  \label{ros1}
\end{equation}
for $t\gg \tau _{c}$, where $\tau _{c}$ is the time within $\sum_{k}\alpha
_{ik}\alpha _{jk}^{\ast }e^{i\omega _{k}\tau }$ have appreciable values. The
constants $k_{ij}$ and $\Delta _{ij}$ are real, defined by 
\begin{equation}
k_{ij}+i\Delta _{ij}=\sum_{k}\alpha _{ik}\alpha _{jk}^{\ast }{\int_{0}^{\tau
_{c}}}d\tau e^{i\left( \omega _{k}-\Omega _{j}\right) \tau }.
\label{k_ij+delta_ij}
\end{equation}

Now we differentiate both sides of Eq. (\ref{ros1}) and iterate to get, in
an analogous way as in Ref. \cite{art19}, 
\begin{equation}
\frac{d}{dt}\mathbf{\tilde{\rho}}_{S}\left( t\right)
=\sum_{i,j=1}^{2}\left\{ \left( k_{ij}+i\Delta _{ij}\right) \left( \mathbf{a}%
_{i}\mathbf{\tilde{\rho}}_{S}\left( t\right) \mathbf{a}_{j}^{\mathbf{\dagger 
}}-\mathbf{\tilde{\rho}}_{S}\left( t\right) \mathbf{a}_{j}^{\mathbf{\dagger }%
}\mathbf{a}_{i}\right) e^{i\left( \Omega _{j}-\Omega _{i}\right) t}\right\} +%
\text{h.c.},
\end{equation}
where terms of second order in $k_{ij}+i\Delta _{ij}$, that are of fourth
order in $\tilde{\mathbf{H}}_{int}\left( t\right) $, were not taken into
account. Returning to the Schr\"{o}dinger picture we write the master
equation 
\begin{equation}
\frac{d}{dt}\mathbf{\rho }_{S}\left( t\right) =\mathcal{L}\mathbf{\rho }%
_{S}\left( t\right) ,  \label{eqmestra}
\end{equation}
where 
\begin{eqnarray}
\mathcal{L} &=&k_{11}\left( 2\mathbf{a}_{1}\bullet \mathbf{a}_{1}^{\mathbf{%
\dagger }}-\bullet \mathbf{a}_{1}^{\mathbf{\dagger }}\mathbf{a}_{1}-\mathbf{a%
}_{1}^{\mathbf{\dagger }}\mathbf{a}_{1}\bullet \right) +i\left( \Delta
_{11}-\Omega _{1}\right) \left[ \mathbf{a}_{1}^{\mathbf{\dagger }}\mathbf{a}%
_{1},\bullet \right] \mathcal{+}  \notag \\
&&k_{22}\left( 2\mathbf{a}_{2}\bullet \mathbf{a}_{2}^{\mathbf{\dagger }%
}-\bullet \mathbf{a}_{2}^{\mathbf{\dagger }}\mathbf{a}_{2}-\mathbf{a}_{2}^{%
\mathbf{\dagger }}\mathbf{a}_{2}\bullet \right) +i\left( \Delta _{22}-\Omega
_{2}\right) \left[ \mathbf{a}_{2}^{\mathbf{\dagger }}\mathbf{a}_{2},\bullet %
\right] \mathcal{+}  \notag \\
&&k_{12}\left( \mathbf{a}_{1}\bullet \mathbf{a}_{2}^{\mathbf{\dagger }}+%
\mathbf{a}_{2}\bullet \mathbf{a}_{1}^{\mathbf{\dagger }}-\bullet \mathbf{a}%
_{2}^{\mathbf{\dagger }}\mathbf{a}_{1}-\mathbf{a}_{1}^{\mathbf{\dagger }}%
\mathbf{a}_{2}\bullet \right) +  \notag \\
&&k_{21}\left( \mathbf{a}_{2}\bullet \mathbf{a}_{1}^{\mathbf{\dagger }}+%
\mathbf{a}_{1}\bullet \mathbf{a}_{2}^{\mathbf{\dagger }}-\bullet \mathbf{a}%
_{1}^{\mathbf{\dagger }}\mathbf{a}_{2}-\mathbf{a}_{2}^{\mathbf{\dagger }}%
\mathbf{a}_{1}\bullet \right) +  \notag \\
&&i\left( \frac{\Delta _{12}-\Delta _{21}}{2}\right) \left( \mathbf{a}%
_{1}\bullet \mathbf{a}_{2}^{\mathbf{\dagger }}-\mathbf{a}_{2}\bullet \mathbf{%
a}_{1}^{\mathbf{\dagger }}-\bullet \mathbf{a}_{2}^{\mathbf{\dagger }}\mathbf{%
a}_{1}+\mathbf{a}_{1}^{\mathbf{\dagger }}\mathbf{a}_{2}\bullet \right) + 
\notag \\
&&i\left( \frac{\Delta _{21}-\Delta _{12}}{2}\right) \left( \mathbf{a}%
_{2}\bullet \mathbf{a}_{1}^{\mathbf{\dagger }}-\mathbf{a}_{1}\bullet \mathbf{%
a}_{2}^{\mathbf{\dagger }}-\bullet \mathbf{a}_{1}^{\mathbf{\dagger }}\mathbf{%
a}_{2}+\mathbf{a}_{2}^{\mathbf{\dagger }}\mathbf{a}_{1}\bullet \right) + 
\notag \\
&&i\left( \frac{\Delta _{12}+\Delta _{21}}{2}\right) \left[ \mathbf{a}_{1}^{%
\mathbf{\dagger }}\mathbf{a}_{2}+\mathbf{a}_{2}^{\mathbf{\dagger }}\mathbf{a}%
_{1},\bullet \right]   \label{Liouvilliano}
\end{eqnarray}
is a Liouvillian superoperator (\textit{i.e.}, an operator which acts on
operators). We use the conventional notation for superoperators \cite{art20}: the
dot sign ($\bullet $) indicates the place to be occupied by $\mathbf{\rho }%
_{S}\left( t\right) $, where the superoperator acts.

In Eq. (\ref{Liouvilliano}) the constants $k_{11}$ and $k_{22}$ are
associated to the individual dissipation of the modes $M_{a}$ and $M_{b}$.
The constants $\Delta _{11}$ and $\Delta _{22}$ have unitary effects,
renormalizing the oscillation frequencies $\Omega _{1}$ and $\Omega _{2}$
(Lamb shifts). The coefficients $k_{12}$, $k_{21}$, $\Delta _{12}$ or $%
\Delta _{21}$ are related to a communication channel between the modes
mediated by the environment, with unitary and non unitary effects over the
evolution of the system. In Eq. (\ref{k_ij+delta_ij}) we see that these
terms will be appreciable only if: 1) $\left| \alpha _{1k}\right| $ and $%
\left| \alpha _{2k}\right| $ are both not zero for several values of $k$
(this means that the system's modes interact effectively with the same
reservoir); 2) $\alpha _{1k}\alpha _{2k}^{\ast }$ have phase correlation for
different $k$ (the system's modes interact with the environment in a
microscopic correlated way). Relatively large values of the cross decay terms%
\emph{\ }$k_{12}+i\Delta _{12}$ and $k_{21}+i\Delta _{21}$ are important for
the appearance of DFS (see Sec. IV). The experimental conditions for it will
be discussed in Sec. V.

\section{Comparing theoretical predictions and experimental results}

\begin{figure}[!ht]
\includegraphics[width=16cm]{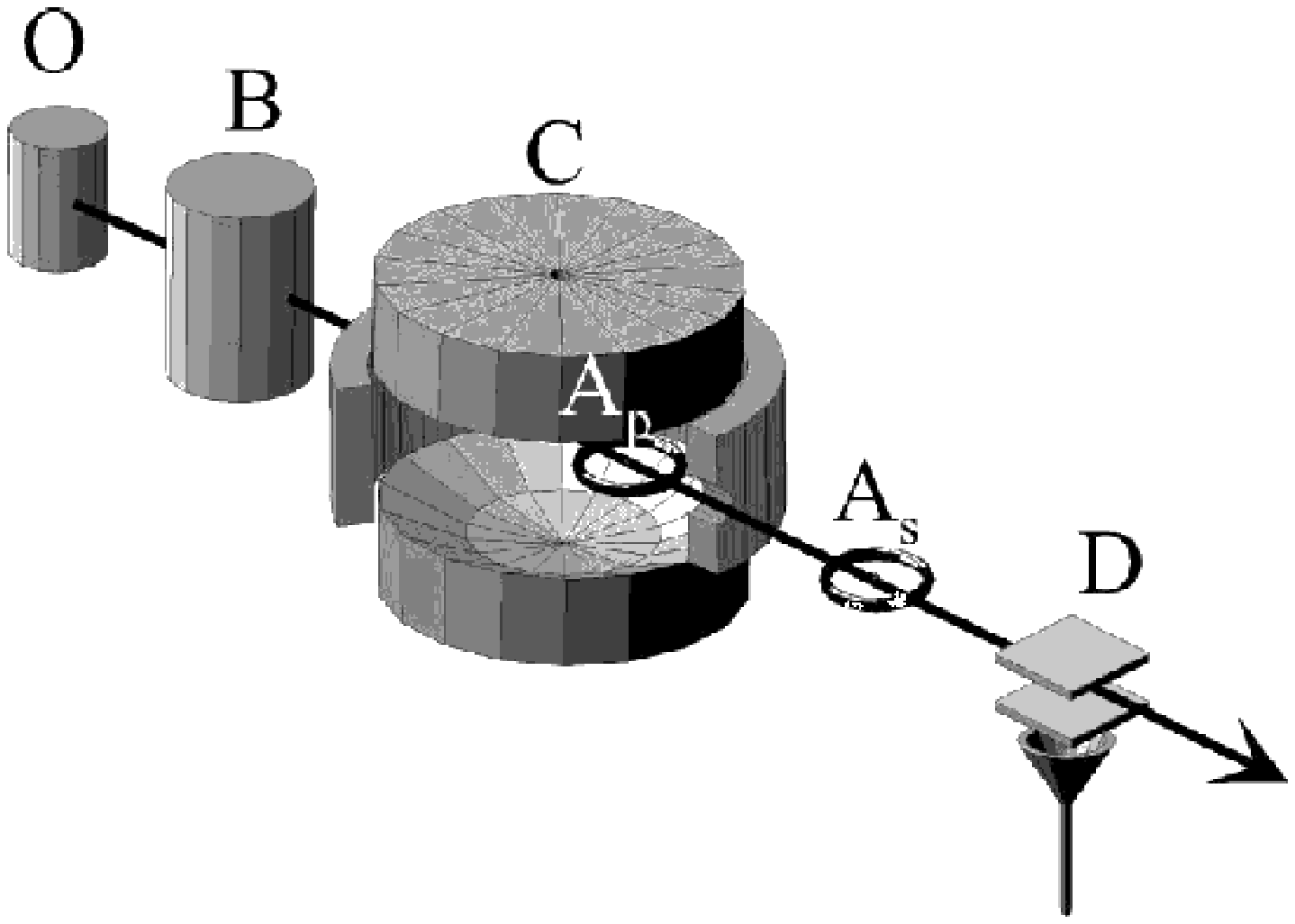}
\vspace{-9cm}
\caption{Sketch of the experiment (figure extracted from Ref.\cite{art14}).}
\label{FigCE1}
\end{figure}

A scheme of the experiment is presented in Fig. \ref{FigCE1}. Cavity \emph{C}
supports two modes, $M_{a}$ and $M_{b}$, with orthogonal polarizations and
different frequencies. The $M_{a}$ mode frequency is larger than that of the 
$M_{b}$ by $\delta $. Circular Rydberg atoms $A_{s}$ and $A_{p}$, with
levels called $e$ and $g$, are sent through the cavity, the first one to
create the entangled state and the second one to reveal its quantum nature.
The detuning $\Delta $ between the $e\longrightarrow g$ transition frequency
and the mode $M_{a}$ may be adjusted by means of the Stark effect. The
sequence of the experiment is described below, closely following Ref. \cite{art14},
where all environmental effects have been ignored.

At time $t=0$, the atom $A_{s}$ enters the cavity in state $\left|
e\right\rangle $. The state of the atom-two modes system is then $\left|
e,0_{a},0_{b}\right\rangle $. The parameter $\Delta $ is initially set to
zero and the atom interacts with $M_{a}$ through a $\frac{\pi }{2}$ Rabi
pulse. The interaction with $M_{b}$ is ignored due to the detuning and the
atom-field coupling is considered to be a constant (vacuum Rabi frequency $%
\Omega $). With an appropriate phase choice of the atomic dipole and
assuming the energy of the state $\left| g,1_{a},0_{b}\right\rangle $ as
zero, the atom-cavity state at $t=\frac{\pi }{2\Omega }$ will be 
\begin{equation*}
\left| \Psi \left( t=\frac{\pi }{2\Omega }\right) \right\rangle =\frac{1}{%
\sqrt{2}}\left[ \left| e_{s},0_{a}\right\rangle +\left|
g_{s},1_{a}\right\rangle \right] \left| 0_{b}\right\rangle .
\end{equation*}
The atom is next, by Stark effect, in resonance with mode $M_{b}$ ($\Delta
=-\delta $). If one now neglects the interaction with mode $M_{a}$ and takes
care of the phases appropriately, one gets for $t=\frac{3\pi }{2\Omega }$%
\begin{equation}
\left| \Psi \left( t=\frac{3\pi }{2\Omega }\right) \right\rangle =\frac{1}{%
\sqrt{2}}\left[ e^{i\phi }\left| 0_{a},1_{b}\right\rangle +\left|
1_{a},0_{b}\right\rangle \right] ,  \label{|psi(t=3pi/2omega)>}
\end{equation}
where $\phi =\frac{\pi }{2}+\frac{\pi \delta }{\Omega }$. The state ($\left|
g\right\rangle $) of the atom $A_{s}$ ends up factorized and need not be
considered any longer.

At time $T$ ($T>\frac{3\pi }{2\Omega }$), atom $A_{p}$ enters the cavity,
whose state will be given by 
\begin{equation*}
\left| \Psi \left( t=T\right) \right\rangle =\frac{1}{\sqrt{2}}\left[ i\exp
\left( \frac{-i\pi \delta }{2\Omega }\right) \exp \left( i\delta T\right)
\left| 0_{a},1_{b}\right\rangle +\left| 1_{a},0_{b}\right\rangle \right] .
\end{equation*}
$A_{p}$ is initially in the ground state and interacts through a $\pi $ Rabi
pulse with $M_{a}$. Next, it interacts with mode $M_{b}$ through a $\frac{%
\pi }{2}$ Rabi pulse, yielding the state 
\begin{equation*}
\left| \Psi \left( t=T+\frac{3\pi }{2\Omega }\right) \right\rangle =\frac{1}{%
2}\left[ i\left| g_{p},1_{b}\right\rangle \left( 1-e^{i\delta T}e^{i\delta
\pi /2\Omega }\right) +\left| e_{p},0_{b}\right\rangle \left( 1+e^{i\delta
T}e^{i\delta \pi /2\Omega }\right) \right] .
\end{equation*}
Mode $M_{a}$ ends up in the vacuum state, factorized. The probability $%
P_{e}(T)$ of finding $A_{p}$ in state $\left| e\right\rangle $ is therefore
given by 
\begin{equation}
P_{e}\left( T\right) =\frac{\left[ 1+\cos \left( \delta T+\Phi \right) %
\right] }{2},  \label{Pe(T)}
\end{equation}
where 
\begin{equation*}
\Phi =\frac{\delta \pi }{2\Omega }.
\end{equation*}

In Appendix A we present a general solution of the Master Equation (\ref
{eqmestra}). This may be used to describe the experiment in question taking
dissipation into consideration in the period between the crossing of the two
atoms. In fact, assuming that in $t=\frac{3\pi }{2\Omega }$ the cavity state
is given by Eq. (\ref{|psi(t=3pi/2omega)>}), in $t=T$, considering the
interaction with the environment, will be 
\begin{eqnarray*}
\rho _{S}\left( T\right)  &=&\frac{1}{2}\left[ \left( e^{i\phi }F_{2}\left(
\tau \right) +L_{2}\left( \tau \right) \right) \left| 0\right\rangle \left|
1\right\rangle +\left( F_{1}\left( \tau \right) +e^{i\phi }L_{1}\left( \tau
\right) \right) \left| 1\right\rangle \left| 0\right\rangle \right] \left[ 
\text{h. c.}\right]  \\
&&+\left( 1-\frac{\left| e^{i\phi }F_{2}\left( \tau \right) +L_{2}\left(
\tau \right) \right| ^{2}}{2}-\frac{\left| F_{1}\left( \tau \right)
+e^{i\phi }L_{1}\left( \tau \right) \right| ^{2}}{2}\right) \left|
0\right\rangle \left| 0\right\rangle \left\langle 0\right| \left\langle
0\right| ,
\end{eqnarray*}
where $\tau =T-\frac{3\pi }{2\Omega }$ and 
\begin{eqnarray}
F_{1}\left( t\right)  &=&\frac{1}{2}\left[ \left( 1-\frac{c}{r}\right)
e^{\lambda _{-}t}+\left( 1+\frac{c}{r}\right) e^{\lambda _{+}t}\right] ,
\label{FiLi} \\
F_{2}\left( t\right)  &=&\frac{1}{2}\left[ \left( 1+\frac{c}{r}\right)
e^{\lambda _{-}t}+\left( 1-\frac{c}{r}\right) e^{\lambda _{+}t}\right] , 
\notag \\
L_{1}\left( t\right)  &=&\frac{1}{2}\left( \frac{k_{12}-i\Delta _{12}}{r}%
\right) \left( e^{\lambda _{-}t}-e^{\lambda _{+}t}\right) ,  \notag \\
L_{2}\left( t\right)  &=&\frac{1}{2}\left( \frac{k_{21}-i\Delta _{21}}{r}%
\right) \left( e^{\lambda _{-}t}-e^{\lambda _{+}t}\right) ,  \notag
\end{eqnarray}
\begin{eqnarray*}
\text{\ \ \ \ \ }\lambda _{-} &=&-R-r,\text{ \ \ \ \ \ \ \ \ \ \ \ \ \ \ \ \ 
}\lambda _{+}=-R+r, \\
c &=&\frac{k_{22}-k_{11}}{2}+i\frac{\left( \Omega _{2}-\Delta _{22}\right)
-\left( \Omega _{1}-\Delta _{11}\right) }{2}, \\
r &=&\sqrt{c^{2}+\left( k_{12}-i\Delta _{12}\right) \left( k_{21}-i\Delta
_{21}\right) }, \\
R &=&\frac{k_{11}+k_{22}}{2}+i\frac{\left( \Omega _{1}-\Delta _{11}\right)
+\left( \Omega _{2}-\Delta _{22}\right) }{2}.
\end{eqnarray*}
Using the above solution for $\rho _{s}(T)$ it is a simple matter to
evaluate $P_{e}(T)$. We get 
\begin{equation*}
P_{e}\left( T\right) =\frac{1}{4}\left| -\left( F_{1}\left( \tau \right)
+e^{i\phi }L_{1}\left( \tau \right) \right) +ie^{2i\Phi }\left( e^{i\phi
}F_{2}\left( \tau \right) +L_{2}\left( \tau \right) \right) \right| ^{2}.
\end{equation*}
The maximum possible values for $\left| k_{12}\right| $, $\left|
k_{21}\right| $, $\left| \Delta _{12}\right| $ and $\left| \Delta
_{21}\right| $ are of the order of $k_{11}$ and $k_{22}$. Given the large
detuning, the experiment is not sensitive to these cross decay constants and
is consistent with $k_{12}=k_{21}=\Delta _{12}=\Delta _{21}=0$. The
experimental results are also consistent with $\Delta _{11}=\Delta _{22}=0$.
In this case we get 
\begin{equation}
P_{e}\left( T\right) =\frac{1}{2}\left[ \frac{e^{-2k_{11}\tau
}+e^{-2k_{22}\tau }}{2}+e^{-\left( k_{11}+k_{22}\right) \tau }\cos \left(
\delta T+\Phi \right) \right] .  \label{Pe(T)1}
\end{equation}
Making $k_{11}=k_{22}=0$, we recover expression (\ref{Pe(T)}).

Note that Eq. (\ref{eqmestra}) has been derived for zero temperature. This
is not the exact experimental condition since in thermal equilibrium the
modes $M_{a}$ and $M_{b}$ contain a small fraction of thermal photons (%
\symbol{126}1). In order to take this effect into account we use effective
dissipation constants for the modes $k_{11}^{ef}=\frac{\bar{n}+1}{2T_{r,a}}$
and $k_{22}^{ef}=\frac{\bar{n}+1}{2T_{r,b}}$ where $\bar{n}$ is the average
number of thermal photons and $T_{r,a}$ ($T_{r,b}$) the measured decay time
for mode $M_{a}$ ($M_{b}$).

There are several sources of imperfection in the experiment, related to the
construction of the modes and to the atomic detection. The fidelity of the
state in Eq. (\ref{|psi(t=3pi/2omega)>}) is estimated to be of $50\%$.
Therefore the interaction with the environment is not the only source of
visibility loss. In order to compare our model results to the experiment we
use a reduction factor of $50\%$.

\begin{figure}[!ht]
\includegraphics[width=16cm]{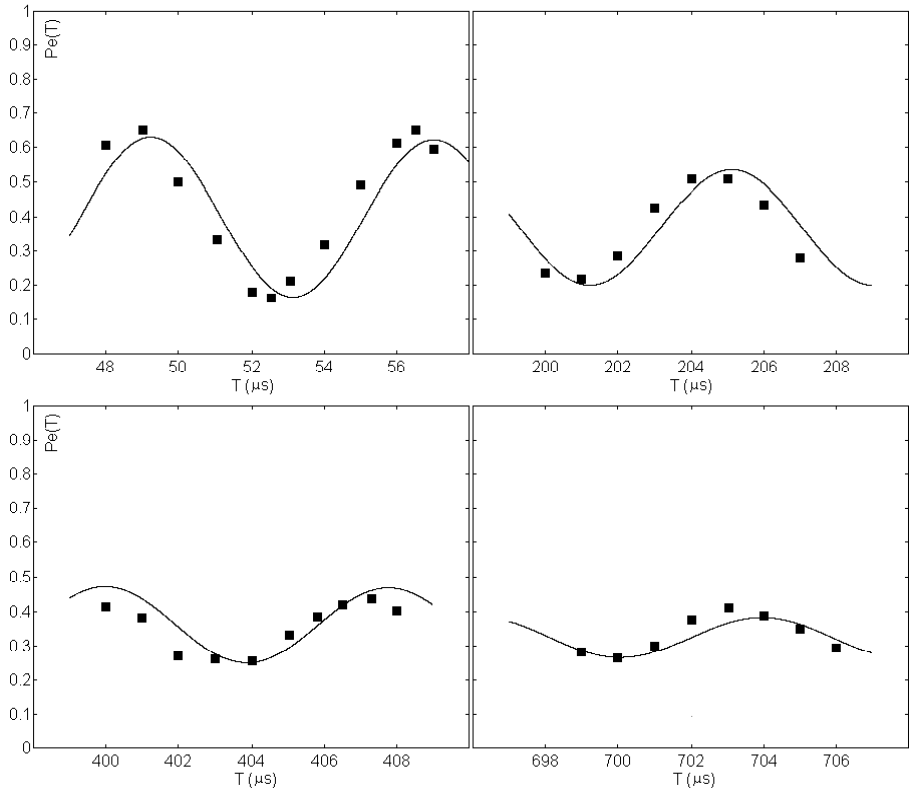}
\vspace{-4cm}
\caption{Probability $P_{e}(T)$ for
detecting $A_{p}$ in state $e$ for the experimental situation described
in Ref. \cite{art14}. The curves refer to the dissipative model. The dots
are from the experiment.}
\label{FigCE2}
\end{figure}

In Fig. \ref{FigCE2} we show $P_{e}(T)\times T$ for the same time windows as
in Ref. \cite{art14}. Eq. (\ref{Pe(T)1}) have been used and the experimental
reduction factor was taken into account. There is good agreement between our
predictions and experiment, especially in what concerns the amplitude and
period of oscillation. We note a slight shift, approximately constant in all
cases ($<1\mu s$), between theory and experiment. Some possible sources of
this phase shift are: 1) The assumption of a constant atom-field
interaction. 2) During the time of switching the atoms off resonance with
mode $M_{a}$ and in resonance with mode $M_{b}$, the phase accumulation
happens in a way that depends on the details of the process, not included in
this model. Notice that these times are, accordingly to \cite{art14}, within $1\mu s$%
, \emph{ie}, in the same order of the shift. 3) We have assumed that the
atom interacts with just one mode at a time. The simultaneous interaction of
the atom with both modes create a communication channel between the modes,
which has consequences on the phase in $P_{e}(T)$. This effect may be more
relevant during the switching of the atom, when the frequency of the atomic
transition is not maximally far from the frequencies of $M_{a}$ and $M_{b}$.

\section{Searching for decoherence-free subspaces}

\begin{figure}[!ht]
\includegraphics[width=16cm]{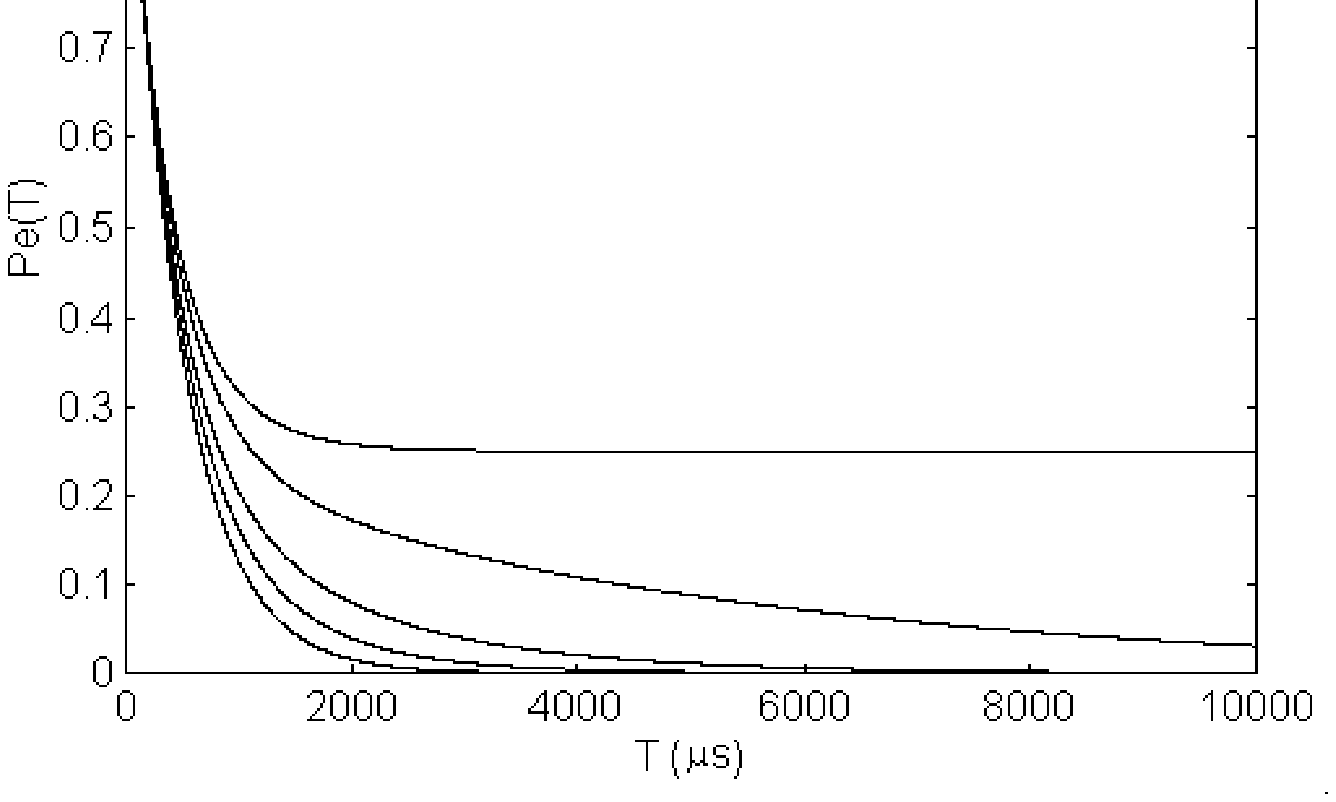}
\vspace{-10cm}
\caption{Probability $P_{e}(T)$ for
detecting $A_{p}$ in state $e$ for resonant modes, according to the
dissipative model. We consider several values for
$k_{12}=k_{21}$: $k_{12}=k_{21}=0$, $k_{12}=k_{21}=0.5\protect\sqrt{k_{11}k_{22}}$, $%
k_{12}=k_{21}=0.7\protect\sqrt{k_{11}k_{22}}$, $k_{12}=k_{21}=0.9\protect%
\sqrt{k_{11}k_{22}}$, $k_{12}=k_{21}=\protect\sqrt{k_{11}k_{22}}$. Upper
curves correspond to higher values for $k_{12}=k_{21}$.}
\label{FigCE3}
\end{figure}

We next perform a mathematical analysis of a different situation than the
experimental one: let us consider resonating modes. In this case $\delta =0$
and the role of $k_{12}$ and $k_{21}$ become significant. In Fig. \ref
{FigCE3} we plot $P_{e}(T)\times T$ for the case, $\delta =0$, $k_{12}=k_{21}
$ for various values of $k_{12}$, keeping the other parameters as considered
in Sec. III. Notice in Fig. \ref{FigCE3} that if $k_{12}=\sqrt{k_{11}k_{22}}$
the field in \emph{C} \emph{does not go to zero} for long times, suggesting
a decoherence-free situation. In fact in this situation the constant $%
\lambda _{-}$ (Eq. (\ref{FiLi})) has a vanishing real part and the
exponentials related to it will therefore not decay. The field decay is then
completely dictated by $\lambda _{+}$. So part of the field remains
protected.

What does this condition $k_{12}=\sqrt{k_{11}k_{22}}$ mean physically? It
means that the environment acts as a\ coherence feed-back mechanism. It
means that photons would scatter from one mode and be transferred to the
other \emph{without loosing their coherence}. This may seem rather
unrealistic although \ it is a sound mathematical consequence of the
extension of a model which works very well for a single mode \cite{art15}. Moreover,
the other curves in Fig. \ref{FigCE3} indicate that even very slight
deviations from this condition already destroy the existence of this
decoherence-free situation.

The experimental scheme to investigate the possibility of such an effect
would need one as small as possible detuning. Of course the assembly must be
altered if we don't want the atom interacting simultaneously with both
modes. We have at least two strategies:

1) In \cite{art14} a quadratic Stark effect is obtained by applying a dc voltage
across the mirrors, which maintains the atomic orbital plane perpendicular
to the cavity axis. In this case the atoms couple equally to both modes. We
may get the Stark effect by, instead, applying two dc voltages perpendicular
to the cavity axis (and perpendicular to each other). Thus the atomic
orbital plane may be maintained perpendicular to the $M_{a}$ or $M_{b}$
polarizations, and the atom may be coupled to just one mode at a time. Such
voltages may be produced directly in the ring around\ the cavity (in this
case the ring must not be continuous) or, if we take away the ring, in
plates outside the cavity.

2) The modes may be constructed in two separate cavities. Since we are
investigating effects of the interaction between the cavities modes through
the environment, it would be suitable to have the cavities as close as
possible to each other. In fact, if the distance between the cavities is
small in the modes' wavelength scale, this interaction is expected to be
maximized.

Adjusting the curve $P_{e}(T)$ could teach us something about the values of $%
k_{12}$ and $k_{21}$, or at least about a tendency to form a DFS, if one
approaches the ideal limit described above.

From the theoretical point of view, it is an important issue to be able to
select a \emph{specific} DFS. Consider that $M_{a}$ and $M_{b}$ have the
same frequency $\omega $ and 
\begin{eqnarray}
k_{22}+i\Delta _{22} &=&\kappa ^{2}\left( k_{11}+i\Delta _{11}\right) ,
\label{condicoesSLD} \\
k_{12}+i\Delta _{12} &=&k_{21}+i\Delta _{21}=\kappa \left( k_{11}+i\Delta
_{11}\right) ,  \notag
\end{eqnarray}
where $\kappa $ is real (a particular choice of $\kappa $ is related to the
quotient between the quality factors of the modes). For all $\Delta _{ij}=0$
this is the case treated
in this section when $k_{12}=k_{21}=\sqrt{k_{11}k_{22}}$. Defining the
bosonic operators 
\begin{eqnarray*}
\mathbf{A} &=&\frac{1}{\sqrt{1+\kappa ^{2}}}\left( \mathbf{a}_{1}+\kappa 
\mathbf{a}_{2}\right) , \\
\mathbf{B} &=&\frac{1}{\sqrt{1+\kappa ^{2}}}\left( \mathbf{a}_{2}-\kappa 
\mathbf{a}_{1}\right) ,
\end{eqnarray*}
we write the Liouvillian (\ref{Liouvilliano}) as  
\begin{eqnarray*}
\mathcal{L} &=&\mathcal{L}_{\mathbf{A}}+\mathcal{L}_{\mathbf{B}}, \\
\mathcal{L}_{\mathbf{A}} &=&-\frac{i}{\hbar }\left( \left[ \mathbf{H}_{%
\mathbf{A}},\bullet \right] \right) +\left( 1+\kappa ^{2}\right) k_{aa}\left( 2%
\mathbf{A}\bullet \mathbf{A}^{\dagger }-\mathbf{A}^{\dagger }\mathbf{A}\bullet
-\bullet \mathbf{A}^{\dagger }\mathbf{A}\right) , \\
\mathcal{L}_{\mathbf{B}} &=&-\frac{i}{\hbar }\left( \left[ \mathbf{H}_{%
\mathbf{B}},\bullet \right] \right) , \\
\mathbf{H}_{\mathbf{A}} &=&\hbar \left( \omega -\Delta _{aa}\left( 1+\kappa
^{2}\right) \right) \mathbf{A}^{\dagger }\mathbf{A}, \\
\mathbf{H}_{\mathbf{B}} &=&\hbar \omega \mathbf{B}^{\dagger }\mathbf{B}.
\end{eqnarray*}
Thus, if a the system is in a state which may be written as 
\begin{equation}
\mathbf{\rho }_{S}=\underset{m,n}{\sum }c_{m,n}\left( \mathbf{B}^{\dagger
}\right) ^{n}\left| 0,0\right\rangle \left\langle 0,0\right| \mathbf{B}^{m}+%
\text{h.c.},  \label{DFS}
\end{equation}
it is not affected by the environment. The states (\ref{DFS}) define a DFS.
Relevant examples for Quantum Optics and Quantum Information are the
coherent state 
\begin{equation*}
\mathbf{\rho }_{S}=\left| -\kappa v\right\rangle \left| v\right\rangle
\left( \text{h.c.}\right) ,
\end{equation*}
the superposition of coherent states 
\begin{equation*}
\mathbf{\rho }_{S}=N\left( \left| -\kappa v\right\rangle \left|
v\right\rangle +e^{i\phi }\left| -\kappa w\right\rangle \left|
w\right\rangle \right) \left( \text{h. c.}\right) 
\end{equation*}
and the superposition of Fock states 
\begin{equation*}
\mathbf{\rho }_{S}=\frac{1}{\sqrt{1+\kappa ^{2}}}\left( \left| 0_{\mathbf{a}%
}\right\rangle \left| 1_{\mathbf{b}}\right\rangle -\kappa \left| 1_{\mathbf{a%
}}\right\rangle \left| 0_{\mathbf{b}}\right\rangle \right) .
\end{equation*}

We shall emphasize that since the conditions (\ref{condicoesSLD}) are
related to the characteristics of the environment, they can not be freely
chosen by the experimenter. To understand physically what may lead to
conditions (\ref{condicoesSLD}), 
consider that the coupling constants of $M_{a}$ and $M_{b}$ to the
reservatory modes may be factorized in the form 
\begin{eqnarray}
\alpha _{1k} &=&\alpha _{1}\gamma _{k},  \label{Infa} \\
\alpha _{2k} &=&\alpha _{2}\gamma _{k},  \notag
\end{eqnarray}
where $\alpha_{1} $ and $\alpha_{2} $ are real numbers.
This corresponds to $M_{a}$ and $M_{b}$ interacting with the environment in
a microscopic correlated way, with a possible difference in the intensity of
the interaction \cite{art21}. For ressonant modes, conditions (\ref{Infa})
imply conditions (\ref{condicoesSLD}) (see Eq. (\ref{k_ij+delta_ij})),
and we
get the DFS just described. Another way to relate conditions (\ref{Infa}) to
this DFS is to use them directly in the Hamiltonian (\ref{Hamiltoniano})
\cite{art21}. An experimental sketch to lead to this microscopic correlation
would need modes as close as possible to each other, preferentialy with the
same polarization, as will be discussed in more detail in the next section.
If the correlation achievable in an experiment is not perfect,
$k_{12}+i\Delta _{12} $ and $k_{21}+i\Delta _{21} $ will assume intermediate
values between zero and the ones in (\ref{condicoesSLD}).

\section{Conclusions}

We deduced a master equation for two oscillators in the presence of a common
reservoir and used it to model a Cavity Quantum Electrodynamics experiment
involving two modes constructed in the same cavity. The theoretical results
show good agreement with the experiment. Such a master equation predicts the
existence of DFS if its cross terms (terms involving operators of both
oscillators) have sufficiently large coefficients. Since the experiment
analyzed is not sensitive to these coefficients, we proposed two slightly
modificated experiments which may permit to investigate them.

The model indicates that a DFS may appear if $M_{a}$ and $M_{b}$ modes
interact with the environmental modes in a microscopically correlated way.
Of course it will not be the case for most systems, and it is not a simple
situation to construct. Probably we may have this microscopic correlation at
least partially for modes whose distance in space is small compared to the
wavelength of the most important environmental modes (the ones with
frequencies near the frequencies of $M_{a}$ and $M_{b}$, as may be seen in
Eq. (\ref{k_ij+delta_ij})). If the environment is composed mainly by
electromagnetic modes, the relevant scale is the scale of the wavelengths of 
$M_{a}$ and $M_{b}$.

The experiment in \cite{art14} was performed with modes in the same place in space.
Unfortunately it doesn't guarantee the microscopic correlation, since $M_{a}$
and $M_{b}$ have orthogonal polarization, and then they ``perceive''
different microscopic environments. Although it must be not easy to built,
there is no theoretical impossibility to construct modes close in the scale
of their proper wavelengths.

The model states clearly that it is very difficult to achieve the parameters
necessary to observe a DFS, and Fig. \ref{FigCE3} shows how fast the DFS is
spoilt when we leave the perfect situation. It is a sign that this model is
realistic. But the model also indicates where are the main difficulties and
what may be done to approach the ideal conditions. Due to the Cavity Quantum
Electrodynamics current experimental stage, the sketches we propose may just
investigate the tendency of forming DFS. If this tendency is confirmed, it
shall encourage later developments.

\begin{acknowledgments}
The authors acknowledge many fruitful discussions with J. G. Peixoto de Faria,
 M. O. Terra Cunha and S. P\'{a}dua. Also financial support from the
brazilian agency CNPq.
\end{acknowledgments}

\appendix

\section{}

Although thoroughly derived in Ref. \cite{art9}, we repeat here the
solution to the master equation, with slight modifications, for
completeness. We will use the parameter derivation technique, which allows
one to determine coefficients $\varsigma _{i}$ such that the identity 
\begin{equation}
e^{\left( \gamma _{1}\mathbf{O}_{1}+\gamma _{2}\mathbf{O}_{2}+\cdots +\gamma
_{n}\mathbf{O}_{n}\right) t}=e^{\varsigma _{1}\left( t\right) \mathbf{O}%
_{1}}e^{\varsigma _{2}\left( t\right) \mathbf{O}_{2}}\cdots e^{\varsigma
_{n}\left( t\right) \mathbf{O}_{n}}  \label{exp(L0t)=exp(d1A1)...exp(dnAn)}
\end{equation}
is valid, where the $\mathbf{O}_{i}$'s are superoperators forming a closed
Lie algebra and $t$ is a parameter. The parameter derivation technique
consists of the following procedure:

\begin{enumerate}
\item  {Derive both sides of Eq. (\ref{exp(L0t)=exp(d1A1)...exp(dnAn)}) with
respect to $t$ and get 
\begin{eqnarray}
\left( \sum_{i=1}^{n}\gamma _{i}\mathbf{O}_{i}\right) \exp \left(
\sum_{i=1}^{n}\gamma _{i}\mathbf{O}_{i}t\right)  &=&\dot{\varsigma}%
_{1}\left( t\right) \mathbf{O}_{1}\prod_{i=1}^{n}e^{\varsigma _{i}\left(
t\right) \mathbf{O}_{i}}  \notag \\
&&+\dot{\varsigma}_{2}\left( t\right) e^{\varsigma _{1}\left( t\right) 
\mathbf{O}_{1}}\mathbf{O}_{2}\prod_{i=2}^{n}e^{\varsigma _{i}\left( t\right) 
\mathbf{O}_{i}}  \notag \\
&&+\cdots   \notag \\
&&+\dot{\varsigma}_{n}\left( t\right) \prod_{i=1}^{n-1}e^{\varsigma
_{i}\left( t\right) \mathbf{O}_{i}}\mathbf{O}_{n}e^{\varsigma _{n}\left(
t\right) \mathbf{O}_{n}}.  \label{(d/dt)exp(L0t)}
\end{eqnarray}
}

\item  {\label{step2}Use the similarity transformation 
\begin{equation}
e^{x\mathbf{O}_{j}}\mathbf{O}_{i}e^{-x\mathbf{O}_{j}}=e^{x\left[ \mathbf{O}%
_{j},\bullet \right] }\mathbf{O}_{i}
\end{equation}
and the linear independence of $\left\{ \mathbf{O}_{i}\right\} $ in order to
obtain differential equations for the parameters }$\varsigma ${$_{i}\left(
t\right) $.}

\item  {Solve the $c$-numbers differential equations and obtain the
factorized evolution superoperator written in the right hand side\ of Eq. (%
\ref{exp(L0t)=exp(d1A1)...exp(dnAn)}).}
\end{enumerate}

As an example of step \ref{step2} above, let us take the second term in the
r.h.s. of Eq. (\ref{(d/dt)exp(L0t)}): 
\begin{eqnarray}
e^{\varsigma _{1}\left( t\right) \mathbf{O}_{1}}\mathbf{O}_{2}
&=&e^{\varsigma _{1}\left( t\right) \mathbf{O}_{1}}\mathbf{O}%
_{2}e^{-\varsigma _{1}\left( t\right) \mathbf{O}_{1}}e^{\varsigma _{1}\left(
t\right) \mathbf{O}_{1}}  \notag \\
&=&e^{\varsigma \left( t\right) \left[ \mathbf{O}_{1},\bullet \right] }%
\mathbf{O}_{2}e^{\varsigma _{1}\left( t\right) \mathbf{O}_{1}}  \notag \\
&\equiv &f_{2}\left( \varsigma _{1}\left( t\right) ,\{\mathbf{O}%
_{i}\}\right) e^{\varsigma _{1}\left( t\right) \mathbf{O}_{1}}.
\end{eqnarray}
Analogously one can carry out a similar operation for all the other terms,
define 
\begin{equation}
f_{3}\equiv f_{3}\left( \varsigma _{1}\left( t\right) ,\varsigma _{2}\left(
t\right) ,\{\mathbf{O}_{i}\}\right) ,\cdots ,f_{n}\equiv f_{n}\left(
\varsigma _{1}\left( t\right) ,\varsigma _{2}\left( t\right) ,\cdots
,\varsigma _{n-1}\left( t\right) ,\{\mathbf{O}_{i}\}\right) 
\end{equation}
and write 
\begin{equation}
\left( \sum_{i=1}^{n}\gamma _{i}\mathbf{O}_{i}\right) \exp \left(
\sum_{i=1}^{n}\gamma _{i}\mathbf{O}_{i}t\right) =\left( \dot{\varsigma}_{1}%
\mathbf{O}_{1}+\dot{\varsigma}_{2}f_{2}+\cdots \dot{\varsigma}%
_{n}f_{n}\right) \exp \left( \sum_{i=1}^{n}\gamma _{i}\mathbf{O}_{i}t\right)
.
\end{equation}
Equivalently: 
\begin{equation}
\sum_{i=1}^{n}\gamma _{i}\mathbf{O}_{i}=\dot{\varsigma}_{1}\mathbf{O}_{1}+%
\dot{\varsigma}_{2}f_{2}+\cdots \dot{\varsigma}_{n}f_{n}.
\end{equation}
Due to the linear independence {of $\left\{ \mathbf{O}_{i}\right\} $}, one
obtains a system of coupled differential equations for the $\varsigma _{i}$%
's comparing the coefficients of each $\mathbf{O}_{i}$.

In our case we have 
\begin{eqnarray}
\mathbf{\rho }_{S}\left( t\right)  &=&e^{\mathcal{L}t}\mathbf{\rho }%
_{S}\left( 0\right)   \notag \\
&=&e^{h_{1}\left( t\right) \mathbf{a}_{1}\mathbf{\bullet a}_{1}^{\dagger
}}e^{h_{2}\left( t\right) \mathbf{a}_{2}\mathbf{\bullet a}_{2}^{\dagger
}}e^{z_{l}\left( t\right) \mathbf{a}_{1}\mathbf{\bullet a}_{2}^{\dagger
}}e^{z\left( t\right) \mathbf{a}_{2}\mathbf{\bullet a}_{1}^{\dagger
}}e^{n_{l}\left( t\right) \mathbf{\bullet a}_{1}^{\dagger }\mathbf{a}%
_{2}}e^{n\left( t\right) \mathbf{a}_{2}^{\dagger }\mathbf{a}_{1}\mathbf{%
\bullet }}  \notag \\
&&e^{m_{2}\left( t\right) \mathbf{a}_{2}^{\dagger }\mathbf{a}_{2}\mathbf{%
\bullet }}e^{p_{2}\left( t\right) \mathbf{\bullet a}_{2}^{\dagger }\mathbf{a}%
_{2}}e^{m_{1}\left( t\right) \mathbf{a}_{1}^{\dagger }\mathbf{a}_{1}\mathbf{%
\bullet }}e^{p_{1}\left( t\right) \mathbf{\bullet a}_{1}^{\dagger }\mathbf{a}%
_{1}}e^{q\left( t\right) \mathbf{a}_{1}^{\dagger }\mathbf{a}_{2}\mathbf{%
\bullet }}e^{q_{l}\left( t\right) \mathbf{\bullet a}_{2}^{\dagger }\mathbf{a}%
_{1}}\mathbf{\rho }_{S}\left( 0\right) .
\end{eqnarray}
Using the method just described we get 
\begin{eqnarray}
i\left( \Delta _{11}-\Omega _{1}\right) -k_{11} &=&\dot{m}_{1}\left(
t\right) -n\left( t\right) \dot{q}\left( t\right) e^{m_{1}\left( t\right)
-m_{2}\left( t\right) },  \notag \\
i\left( \Delta _{22}-\Omega _{21}\right) -k_{22} &=&\dot{m}_{2}\left(
t\right) +n\left( t\right) \dot{q}\left( t\right) e^{m_{1}\left( t\right)
-m_{2}\left( t\right) },  \notag \\
i\Delta _{12}-k_{12} &=&\dot{q}\left( t\right) e^{m_{1}\left( t\right)
-m_{2}\left( t\right) },  \notag \\
i\Delta _{21}-k_{21} &=&\dot{n}\left( t\right) +n\left( t\right) \left( \dot{%
m}_{1}\left( t\right) -\dot{m}_{2}\left( t\right) \right) -n\left( t\right)
^{2}\dot{q}\left( t\right) e^{m_{1}\left( t\right) -m_{2}\left( t\right) }, 
\notag \\
i\left( \Omega _{1}-\Delta _{11}\right) -k_{11} &=&\dot{p}_{1}\left(
t\right) -n_{l}\left( t\right) \dot{q}_{l}\left( t\right) e^{p_{1}\left(
t\right) -p_{2}\left( t\right) },  \notag \\
i\left( \Omega _{2}-\Delta _{22}\right) -k_{22} &=&\dot{p}_{2}\left(
t\right) +n_{l}\left( t\right) \dot{q}_{l}\left( t\right) e^{p_{1}\left(
t\right) -p_{2}\left( t\right) },  \notag \\
-i\Delta _{12}-k_{12} &=&\dot{q}_{l}\left( t\right) e^{p_{1}\left( t\right)
-p_{2}\left( t\right) },  \notag \\
-i\Delta _{21}-k_{21} &=&\dot{n}_{l}\left( t\right) +n_{l}\left( t\right)
\left( \dot{p}_{1}\left( t\right) -\dot{p}_{2}\left( t\right) \right)
-n_{l}\left( t\right) ^{2}\dot{q}_{l}\left( t\right) e^{p_{1}\left( t\right)
-p_{2}\left( t\right) },  \notag \\
2k_{11} &=&z\left( t\right) \left( i\Delta _{21}-k_{21}\right)   \notag \\
&&-z_{l}\left( t\right) \left( i\Delta _{21}+k_{21}\right) +h_{1}\left(
t\right) \left( -2k_{11}\right) +\dot{h_{1}}\left( t\right) ,  \notag \\
2k_{22} &=&z_{l}\left( t\right) \left( i\Delta _{12}-k_{12}\right)   \notag
\\
&&-z\left( t\right) \left( i\Delta _{12}+k_{12}\right) +h_{2}\left( t\right)
\left( -2k_{22}\right) +\dot{h_{2}}\left( t\right) ,  \notag \\
i\left( \Delta _{21}-\Delta _{12}\right) +k_{21}+k_{12} &=&z\left( t\right)
\left( i\left( \Omega _{1}-\Delta _{11}-\Omega _{2}+\Delta _{22}\right)
-k_{11}-k_{22}\right)   \notag \\
&&-h_{2}\left( t\right) \left( i\Delta _{21}+k_{21}\right) +h_{1}\left(
t\right) \left( i\Delta _{12}-k_{12}\right) +\dot{z}\left( t\right) ,  \notag
\\
i\left( \Delta _{12}-\Delta _{21}\right) +k_{12}+k_{21} &=&z_{l}\left(
t\right) \left( i\left( \Omega _{2}-\Delta _{22}-\Omega _{1}+\Delta
_{11}\right) -k_{22}-k_{11}\right)   \notag \\
&&-h_{1}\left( t\right) \left( i\Delta _{12}+k_{12}\right) +h_{2}\left(
t\right) \left( i\Delta _{21}-k_{21}\right) +\dot{z_{l}}\left( t\right) .
\end{eqnarray}
The solution reads 
\begin{eqnarray}
n\left( t\right)  &=&\frac{L_{2}\left( t\right) }{F_{1}\left( t\right) }%
,\quad \quad q\left( t\right) =\frac{L_{1}\left( t\right) }{F_{1}\left(
t\right) },  \notag \\
e^{m_{1}\left( t\right) } &=&F_{1}\left( t\right) ,\quad \quad
e^{m_{2}\left( t\right) }=e^{-2Rt}e^{-m_{1}\left( t\right) },  \notag \\
h_{1}\left( t\right)  &=&\left( \left| F_{2}\left( t\right) \right|
^{2}+\left| L_{2}\left( t\right) \right| ^{2}\right) e^{4k_{m}t}-1,  \notag
\\
h_{2}\left( t\right)  &=&\left( \left| F_{1}\left( t\right) \right|
^{2}+\left| L_{1}\left( t\right) \right| ^{2}\right) e^{4k_{m}t}-1,  \notag
\\
z\left( t\right)  &=&-\left( L_{1}\left( t\right) F_{2}^{\ast }\left(
t\right) +L_{2}^{\ast }\left( t\right) F_{1}\left( t\right) \right)
e^{4k_{m}t},  \notag \\
z_{l}\left( t\right)  &=&z^{\ast }\left( t\right) ,  \notag \\
n_{l}\left( t\right)  &=&\left( n\left( t\right) \right) ^{\ast },\quad
\quad q_{l}\left( t\right) =\left( q\left( t\right) \right) ^{\ast },  \notag
\\
p_{2}\left( t\right)  &=&\left( m_{2}\left( t\right) \right) ^{\ast },\quad
\quad p_{1}\left( t\right) =\left( m_{1}\left( t\right) \right) ^{\ast },
\end{eqnarray}
where $F_{1}\left( t\right) $, $F_{2}\left( t\right) $, $L_{1}\left(
t\right) $ and $L_{2}\left( t\right) $ are given by Eqs. (\ref{FiLi}).

\begin{thebibliography}{21}

\bibitem{art1} D. Giulini, E. Joos, C. Kiefer, J. Kupsch, I.-O. Stamatescu, H. D. Zeh, 
\textit{Decoherence and the Appearance of a Classical World in Quantum Theory%
} (Springer-Verlag, Berlin, 1996).

\bibitem{art2} D. Bouwmeester, A. Ekert, and A. Zeilinger, eds, \textit{The Physics of
Quantum Information} (Springer-Verlag, Berlin, 2000).

\bibitem{art3} M. Brune and S. Haroche, in \textit{Quantum Dynamics of simple Systems},
G-L Oppo, S. M. Barnett, E. Riis, M. Wilkinson, P. Osborne, eds. (SUSSP
Publications and Institute of Physics Publishing, 1996).

\bibitem{art4} G. S. Agarwal, \textit{Quantum Statistical Theories of Spontaneous
Emission and their Relation to other Approaches}, edited by G. H\"{o}hler,
Springer Tracts in Modern Physics, Vol. 70 (Springer-Verlag, Berlin, 1974),
Section 15.B.

\bibitem{art5} C. W. Gardiner, \textit{Phys. Rev. Lett.} \textbf{70}, 2269 (1993).

\bibitem{art6} H. J. Carmichael, \textit{Phys. Rev. Lett.} \textbf{70}, 2273 (1993).

\bibitem{art7} C. W. Gardiner, and P. Zoller, \textit{Quantum Noise, 2}nd edition,
Springer, Berlin, 2000, Chapter 12.

\bibitem{art8} Uzma Akram, Z. Ficek, and S. Swain, \textit{Phys. Rev. A} \textbf{62},
013413 (2000).

\bibitem{art9} A. R. Bosco de Magalh\~{a}es, S. G. Mokarzel, M. C. Nemes, and M. O.
Terra Cunha, \textit{Physica A} \textbf{341}, 234 (2004),
also quant-ph/0405022.

\bibitem{art10} P. Kochan, and H. J. Carmichael, \textit{Phys. Rev. A} \textbf{50},
1700 (1994).

\bibitem{art11} C. W. Gardiner, and A. S. Parkins, \textit{Phys. Rev. A} \textbf{50},
1792 (1994).

\bibitem{art12} A. S. Parkins, and H. J. Kimble, \textit{J. Opt. B: Quantum Semiclass.
Opt. }\textbf{1}, 496 (1999).

\bibitem{art13} Z. Ficek and S. Swain, \textit{Jour. of Mod. Opt.} \textbf{\ 49}, 3
(2002).

\bibitem{art14} A. Rauschenbeutel, P. Bertet, S. Osnaghi, G. Nogues, M. Brune, J. M.
Raimond, and S. Haroche, \textit{Phys. Rev. A} \textbf{64}, 050301(R) (2001).

\bibitem{art15} A. O. Caldeira and A. J. Legget, \textit{Annals of Physics} \textbf{149}%
, 374 (1983).

\bibitem{art16} M. Dub\'{e} and P. C. E. Stamp, \textit{Chem. Phys.} 
\textbf{268}, 257 (2001),
also cond-mat/0102156v2.

\bibitem{art17} W. H. Louisell, \textit{Quantum Statistical Properties of Radiation }%
(Wiley, New York, 1973), Section 6.1.

\bibitem{art18} R. Kubo, M. Toda, N. Hashitsumi, \textit{Statistical Physics II -
Nonequilibrium Statistical Physics} (Springer-Verlag, 1978 - Springer Series
in Solid-State Sciences, n. 31).

\bibitem{art19} Juan Pablo Paz, Salman Habib, and Wojciech H. Zurek, (Lectures given by
both authors at the 72nd Les Houches Summer School on ``Coherent Matter
Waves '', 1999), quant-ph/0010011, Section 3.1.\textit{\ }

\bibitem{art20} J. G. Peixoto de Faria, and M. C. Nemes, \textit{Phys. Rev. A} \textbf{%
59}, 3918 (1999).

\bibitem{art21} K. M. Fonseca Romero, S. G. Mokarzel, M. O. Terra Cunha, and M. C.
Nemes, quant-ph/0304018.

\end{thebibliography}
\end{document}